\documentclass[aps,pre,twocolumn,groupedaddress,showpacs,showkeys,amsmath]{revtex4}
\usepackage{graphicx}
\begin{document}
\title{Nodal domains in open microwave systems}
\author{U.~Kuhl}
\affiliation{Fachbereich Physik, Philipps-Universit\"{a}t Marburg,
Renthof 5, D-35032 Marburg, Germany}
\author{R.~H\"{o}hmann}
\affiliation{Fachbereich Physik, Philipps-Universit\"{a}t Marburg,
Renthof 5, D-35032 Marburg, Germany}
\author{H.-J. St\"{o}ckmann}
\affiliation{Fachbereich Physik, Philipps-Universit\"{a}t Marburg,
Renthof 5, D-35032 Marburg, Germany}
\author{S. Gnutzmann}
\affiliation{School of Mathematical Sciences, University of
  Nottingham, University Park, Nottingham, NG7 2RD, United Kingdom}
\affiliation{Fachbereich Physik, Freie Universit\"{a}t Berlin,
  Arnimallee 14, 14195 Berlin, Germany}
\date{\today}
\begin{abstract}
 Nodal domains are studied both for real $\psi_R$ and imaginary part $\psi_I$ of the wavefunctions of an open microwave cavity and found to show the same behavior as wavefunctions in closed billiards. In addition we investigate the variation of the number of nodal domains and the signed area correlation by changing the global phase $\varphi_g$ according to $\psi_R+i\psi_I=e^{i\varphi_g}(\psi_R'+i\psi_I')$. This variation can be qualitatively, and the correlation quantitatively explained in terms of the phase rigidity characterising the openness of the billiard.
\end{abstract}

\maketitle

\section{Introduction}

Approximately 200 years after Chladni's work on acoustic plates detecting and classifying nodal lines \cite{Chl02} there is a revival due to two pioneering works on the subject of nodal domains. The first to mention is by Blum et~al.~\cite{blu02} proposing nodal domains as a tool to separate integrable and chaotic systems from each other. They explained their numerical findings for nodal domains in chaotic billiards in terms of the random plane wave approach \cite{ber77a} and found a good agreement. Bogomolny and coworkers \cite{bog02b} introduced a percolation model, yielding explicit results, e.\,g., for the linear increase of number of nodal domains with the quantum number with a slope of 0.0624. These results have been experimentally verified in closed microwave billiards \cite{Sav04b,Hul05b}.

Other studies of nodal domains are concerned with the continuation into the classically forbidden region taken tunneling into account \cite{bie02}, with graphs \cite{gnu04}, with iso-spectrality on graphs \cite{ban06} or billiards, i.\,e., whether the shape of the drum can be determined by nodal domain counting \cite{gnu05,gnu06,lev06}, with quantum maps \cite{kea06}, and with the anharmonic oscillator \cite{aib05}.

Here we will concentrate on real physical systems, which are always open due to the measurement process and absorption. We shall show that nodal domains still can be defined and have the same behavior as in closed systems. In addition it will be illustrated that the variation of the number of nodal domains in dependence of a global phase of the wavefunction can be used as a tool to determine the `openness' of the billiard.

\section{Nodal domains for real and imaginary parts}

\begin{figure}
  \includegraphics[width=\columnwidth]{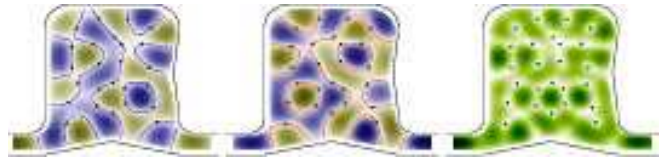}
  \caption{\label{fig:wavefunction} (color online) Real part $\psi_R$ (left), imaginary part $\psi_I$ (center), and modulus$|\psi|^2=\psi_R^2+\psi_I^2$ of the wavefunction $\psi$ at $\nu=5.64$\,GHz corresponding to a Weyl number of $n_{\rm{Weyl}}$=32.5. Correlations between $\psi_R$ and $\psi_I$  have already been removed by a global phase, see text. Additionally the nodal points of the modulus and the nodal lines of real and imaginary parts are marked. Nodal points occur at crossings of a nodal lines of $\psi_R$ with that of $\psi_I$.}
\end{figure}

The billiard used is a quantum dot like structure with two attached leads of width 2\,cm (for more details see \cite{kim02}). Additionally we added two insets to two of the sides (see also Fig.~\ref{fig:wavefunction}) to avoid any bouncing ball structures \cite{kim02,kim03a}. The measurements have been performed on a grid of step size 2.5\,mm. A similar billiard has already been used to determine long range correlations in the wavefunctions \cite{kim05a} and vortex distributions and correlations \cite{kim03b}. For each point the transmission $S_{12}$ from a fixed antenna 1 in one lead to the probing antenna 2 has been measured including the phases. As the transmission depends only on the field distribution at the positions of the antennas, and the position of antenna 1 is fixed, one can obtain the electric field distribution $E_z(x,y)$ at the position of antenna 2 \cite{ste95,stoe99}. We will use the term `wavefunction' for this quantity as it is related to the quantum mechanical problem of a particle in a box, even though it is not the eigenfunction of a single resonance but a superposition of different eigenfunctions. From this measurement the wavefunction $\psi(x,y)=E_z(x,y)$ has been determined including its real and imaginary parts (see \cite{ste95,kuh05b} and chapter 6 of \cite{stoe99}). In Fig.~\ref{fig:wavefunction} real part $\psi_R$, imaginary part $\psi_I$ and the modulus of a typical wavefunction is shown.

\begin{figure}
  \includegraphics[width=\columnwidth]{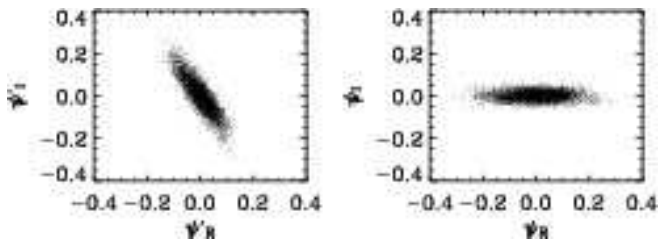}
  \caption{\label{fig:ReIm} Imaginary versus real part of the wave
    function at $\nu=13.84$\,GHz, for the directly measured wave
    function (left), and after uncorrelating them (right) by
    $\psi_R+i\psi_I=e^{-i\varphi_{g,0}} \left(\psi_R'+i\psi_I' \right)$. $\varphi_{g,0}$ is a global phase from the experiment acquired due to the antenna and leads.}
\end{figure}

Fig.~\ref{fig:ReIm}(left) shows a plot of the imaginary part $\psi_I'$ versus the real part $\psi_R'$ for a typical wavefunction prior to any correction. Each dot corresponds to a measured grid point. Usually the cloud of dots is not aligned with its main axes with respect to the coordinate axes leading to a correlation of real and imaginary part. This correlation has its origin in unwanted global phase shifts $\varphi_{g,0}$ mainly from the lead and the antennas. This global phase has been removed together with the correlations between $\psi_R$ and $\psi_I$ by means of a phase rotation,
\begin{equation}\label{eq:ExpGlobalPhase}
  \psi_R+i\psi_I=e^{-i\varphi_{g,0}} \left(\psi_R'+i\psi_I' \right)
\end{equation}
where $\langle\psi_R^2\rangle > \langle\psi_I^2\rangle$, thus adjusting the main axes of the cloud of dots to the coordinate axes (Fig.~\ref{fig:ReIm} (right)). For a completely open system a circular cloud would have been expected. The eccentricity thus reflects the lack in openness (for this wavefunction). For a quantitative description of the openness the phase rigidity $|\rho|^2$ has been introduced \cite{lan97}. $\rho$ is defined by
\begin{equation}\label{eq:rigidity}
  \rho=\int d\mathbf{r}\,\psi(\mathbf{r})^2 =
  \frac{\langle\psi_R^2\rangle-\langle\psi_I^2\rangle}
  {\langle\psi_R^2\rangle+\langle\psi_I^2\rangle}.
\end{equation}
This quantity is immediately available from the cloud of dots as shown in (Fig.~\ref{fig:ReIm}(right)). It is noteworthy to mention that the phase rigidity is a highly fluctuating quantity and is different for each wavefunction.

Because of the openness of the billiard there is a strong overlap of eigenfrequencies at each frequency making a direct determination of the mode number $n$ impossible. Therefore the Weyl number \cite{wey13} has been used to determine this quantity, in our billiards given by
\begin{equation}\label{eq:Weyl}
  n_{\rm{Weyl}}=\frac{A}{4\pi} k^2 - \frac{S}{4\pi} k + C
\end{equation}
where $k=2\pi\nu/c$ is the wavenumber, $\nu$ the frequency, $c$ the speed of light, $A$ and $S$ are the area and the circumference of the billiard and $C$ is a constant determined by corners and the curvature in the billiards. The order of the constant is 1 and in our case we used $A \approx$\,0.0361\,m$^2$, $S\approx$\,0.807\,m and $C$=0. Strictly speaking the Weyl formula is not applicable for open systems, since the resonances are shifted from the real axis into the complex plane. Occasionally resonances even may be removed from the spectrum due to a strong coupling to the surroundings \cite{leh95}, and fractal Weyl laws may occur \cite{lu03,non}. But still the Weyl formula is the best approximation known in this case to obtain the mode number $n$.

From the measured real and imaginary parts we determined the nodal lines by means of a bilinear interpolation. From the bilinear interpolation we get the lines within the grid. Using this information and the Hoshen-Kopelman method \cite{hos76} we obtain the number of nodal domains $\nu_n$ and their areas.

\begin{figure}
  \includegraphics[width=8cm]{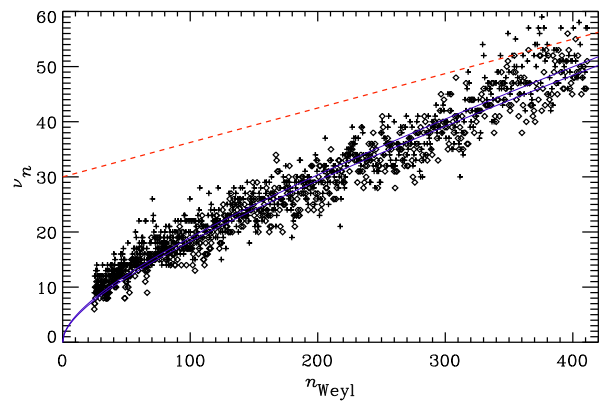}\\
  \caption{\label{fig:nnodal} (color online) The number of nodal
    domains $\nu_n$ versus the Weyl number $n_{\rm{Weyl}}$ for the real
    (crosses) and imaginary part (diamonds). The red dashed line
    correspond to the theoretical prediction $\nu_n=0.0624 n$
    of the percolation model
    \cite{bog02b}. The blue solid lines are fits including boundary
    effects \cite{blu02}
    with Eq.~(\ref{eq:NodalDomainNumber}) with
    a linear slope of 0.059 and 0.060 for real and imaginary
    part respectively.}
\end{figure}

In the case of open systems the distribution of intensities $P(|\psi|^2)$, e.\,g., can be obtained from the random superposition of plane waves (RSPW) approach \cite{ber77a}, expressing $\psi$ as a
\begin{equation}
  \psi(\mathbf{r}) = \sum_{\mathbf{k}} a(\mathbf{k})
  e^{i \mathbf{k} \cdot \mathbf{r}}.
  \label{eq:sum}
\end{equation}
In open systems the plane wave amplitudes $a(\mathbf{k})$ have a
Gaussian distribution with zero mean and with variance
\begin{equation}
  \langle a(\mathbf{k}) a(-\mathbf{k}) \rangle =  \rho \langle a(\mathbf{k}) a^*(\mathbf{k}) \rangle,
  \label{eq:avar}
\end{equation}
where $|\rho|^2$ is the phase rigidity as given by Eq.~(\ref{eq:rigidity}). In this model the phase rigidity is only defined by an average and individual realizations therefore have fluctuating phase rigidities, depending on the number of plane waves used.

First we investigate the number of nodal domains $\nu_n$ for the real and imaginary part. Bogomolny and Schmidt \cite{bog02b} predicted from the percolation model a linear increase with a slope of $a$=0.0624. Blum et~al.\ \cite{blu02} introduced an additional term to take boundary effects into account,
\begin{equation}
  \label{eq:NodalDomainNumber}
  \nu_n= a n + b \sqrt{n}
\end{equation}
where $n$ is the quantum number, which in our case is given by the Weyl number $n_{\rm{Weyl}}$ (see Eq.~(\ref{eq:Weyl})). In Fig.~\ref{fig:nnodal} number of nodal domains is shown versus the Weyl number for the real (crosses) and imaginary (diamonds) part of the wavefunction. The dashed line corresponds to the slope given by Ref.~\onlinecite{bog02b}. The data were fitted with Eq.~(\ref{eq:NodalDomainNumber}) and we found $a$=0.0594 and $b$=1.231 for the real and $a$=0.0599 and $b$=1.300 for the imaginary part. The value for $a$ is in accordance with the expected slope of 0.0624.

The boundary correction parameter $b$ should scale with the ratio $S/\sqrt{A}$, where $S$ and $A$ are the circumference and area of the billiard, respectively. The rescaled quantity $b_c=b\sqrt{A}/S$ should be universal. In the present case we find values of 0.29 and 0.31 for $b_c$ for the real and the imaginary part, respectively. For two other billiards investigated experimentally,namely the half rough circle billiard with \cite{Hul05b} and without \cite{Sav04b} teflon insert, values of 0.27, and 0.20 have been obtained for $b_c$. The $b_c$ obtained for the billiard with a half circle teflon insert have to be taken with care, as the internal boundary between teflon and air and the integrable shape of the teflon insert may give rise to deviations. In case of the numerical calculations presented in Ref.~\onlinecite{blu02} values of 0.25 and 0.21 have been obtained for the Sinai and stadium billiard, respectively. In all cases the estimated error is of about 10\%. Thus $b_c$ is comparable for all billiards.

\begin{figure}
  \includegraphics[width=8cm]{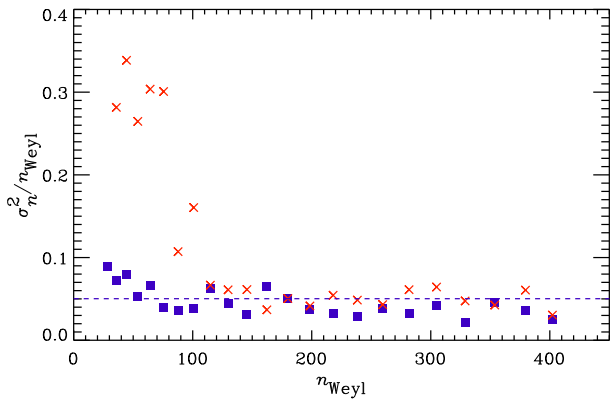}\\
  \caption{\label{fig:variance} (color online) The variance of the
    number of nodal domains $\nu_n$ versus the Weyl number
    $n_{\rm{Weyl}}$ for the real (blue boxes) and imaginary part (red
    crosses). The blue dashed line corresponds to the theoretical
    prediction.}
\end{figure}

For the variance $\sigma^2$ of the number of nodal domains the percolation model predicts $\sigma^2\approx 0.05 n$ \cite{bog02b}. In Fig.~\ref{fig:variance} the scaled variance $\sigma^2/n_{\rm{Weyl}}$ is plotted versus the Weyl number $n_{\rm{Weyl}}$ for the real (blue boxes) and imaginary (red crosses) part. The variances have been calculated from the corresponding fits mentioned before and the average has been performed over 30 consecutive wavefunctions. A good agreement is found. The larger variations at small Weyl numbers for the imaginary part are due to experimental noise: As in this regime the phase rigidity is often small, i.\,e., the imaginary part is small, noise in the experimental data can create small nodal domains particularly close to the billiard boundary, leading to higher fluctuations.

\begin{figure}
  \includegraphics[width=8cm]{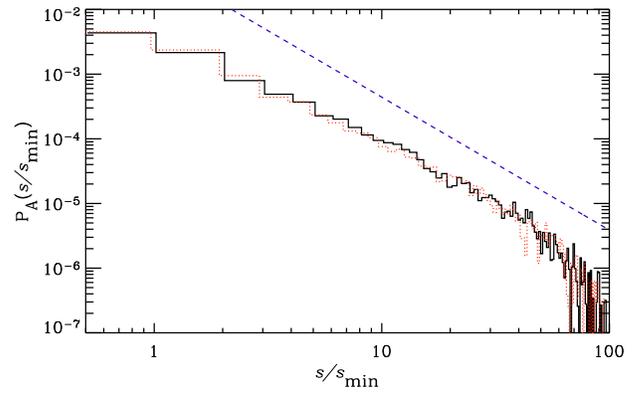}\\
  \caption{\label{fig:PAreas} (color online) The distribution of normalized nodal domain areas $P_A(s/s_{\rm{min}})$ is shown for the real (solid histogram) and the imaginary part (red dotted histogram). Additionally the expected algebraic decay with an exponent of $\tau=187/91$ is shown as blue dashed line.}
\end{figure}

Another quantity of interest is the distribution $P_A$ of nodal domain areas $s$. The area distribution has also been calculated in Ref.~\onlinecite{bog02b} and is given asymptotically by
\begin{equation}\label{eq:PArea}
  P_A(s)\propto \left(\frac{s}{s_{\rm{min}}}\right)^{-\tau}
\end{equation}
where $\tau=187/91$ and $s_{\rm{min}}=\pi (j_1/k)^2$, the smallest possible area for fixed wavenumber $k$, where $j_1$ is the first zero of the Bessel function $J_0(x)$. In Fig.~\ref{fig:PAreas} the distribution of normalized nodal domain areas is shown for the real and imaginary part. All evaluated wavefunctions have been used to obtain the distribution. Additionally the expected slope of the algebraic decay with $\tau=187/91$ is shown as a dashed line and is in agreement with the experimental histograms. Deviations at large values of $s/s_{\rm{min}} > $ 80 are due to poor statistics. Here only a few entries per bin are available.

No difference between the expected behavior for closed chaotic systems and the behavior for the real and imaginary part of the wavefunction in open chaotic systems is found. A global phase rotation had been performed to uncorrelate real and imaginary parts, but the same behavior is found for the number of nodal domains (Fig.~\ref{fig:nnodal}), its variance (Fig.~\ref{fig:variance}) and the area distribution (Fig.~\ref{fig:PAreas}), using the original data without any global phase rotation.

\section{Nodal domain dependence on global phase}

\begin{figure*}
  \includegraphics[width=2\columnwidth]{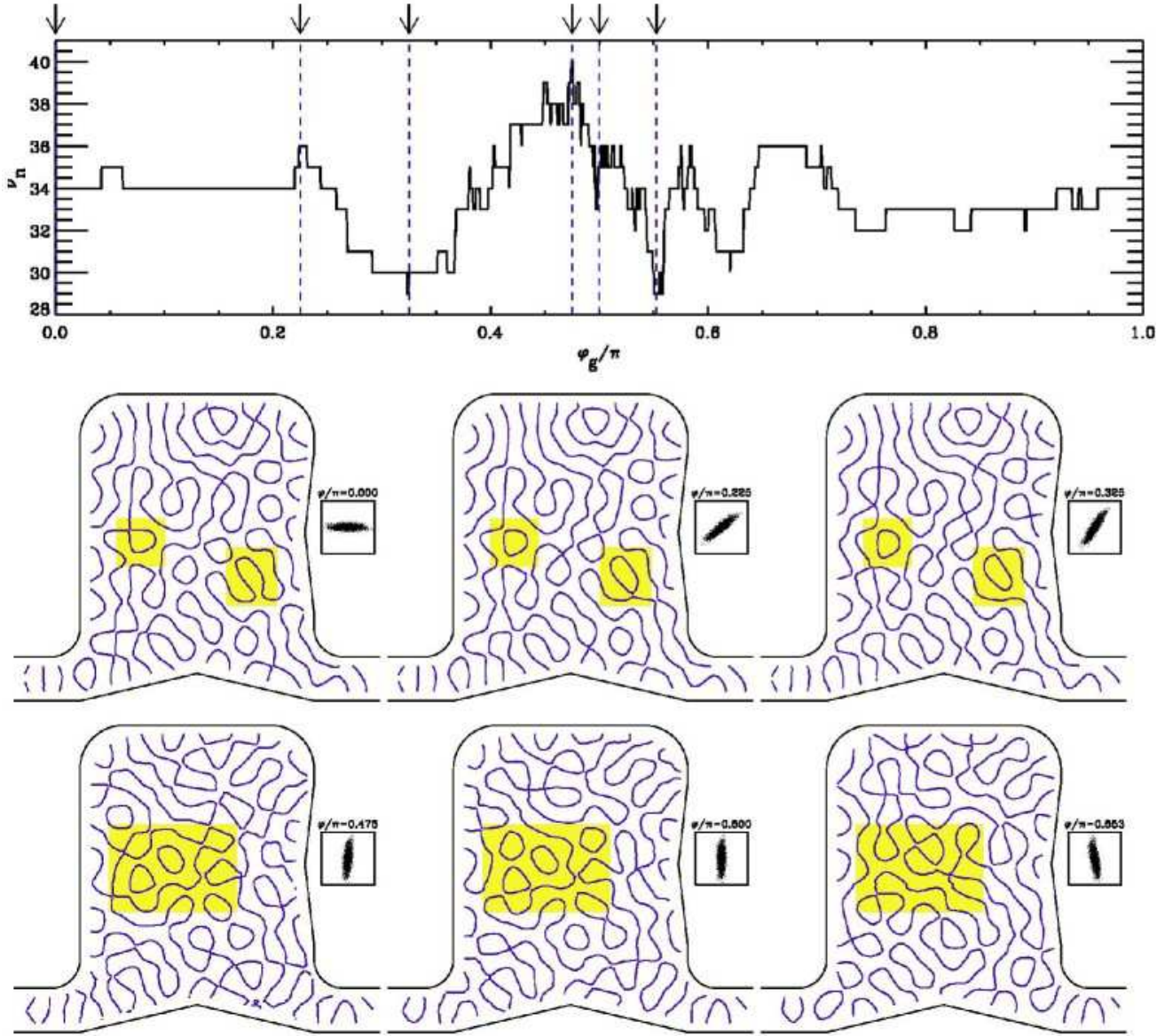}
  \caption{\label{fig:NodalPhi} (color online) (Upper part) Number of nodal domains $\nu_n$ as a function of a global phase $\varphi_g$ for a wavefunction at $n_{\rm{Weyl}} \approx$ 223. Marked are the phases $\varphi_g$, for which the nodal domains are shown below, corresponding to $\varphi_g/\pi$ = 0, 0.225, 0.35, 0.475, 0.5, and 0.5525. Areas, where appearances and disappearances of nodal domains can be seen, are highlighted. The inserts show the corresponding $\psi_I$ versus $\psi_R$ plots, see also Fig.~\ref{fig:ReIm}.}
\end{figure*}

In the preceding section we have mostly neglected the fact that the billiard is open and treated the real and imaginary part just like a real valued wavefunction in case of closed systems. But as the billiard is open we have additional parameters like the rigidity $|\rho^2|$ or global phases $\varphi_g$ or even new quantities like currents $j$ or vorticities $\omega$ \cite{bar02,kim03a}. In Fig.~\ref{fig:NodalPhi} we present the nodal domains and the number of nodal domains of the real part of a wavefunction at Weyl number $n_{\rm{Weyl}} \approx$ 223 with a phase rigidity $|\rho|^2 \approx 0.81$. At the top the number of nodal domains for the real part is shown as a function of the global phase $\varphi_g$. For $\varphi_g=0$ real and imaginary part are uncorrelated and for $\varphi_g=\pi/2$ real and imaginary part have changed their identities and are uncorrelated again. In the lower part a selection of nodal domains is presented for phases indicated by arrows in the upper figure. Additionally the $\psi_I$ versus $\psi_R$ plots are shown, see Fig.~\ref{fig:ReIm}. Highlighted rectangles emphasize regions, where changes of the number of nodal domains occur, due to rearrangements of nodal lines. While the phase is changing the nodal lines are shifted and permanently dissolved and reconnected. The only points fixed are the nodal points of the corresponding modulus $|\psi|^2$, corresponding to elliptic points in the flow. Nodal lines can only intersect at the saddle points corresponding to hyperbolic points in the flow. Their position is also independent of the global phase (see Ref.~\cite{hoeh} for a more detailed discussion of these features).

\begin{figure}
  \includegraphics[width=8cm]{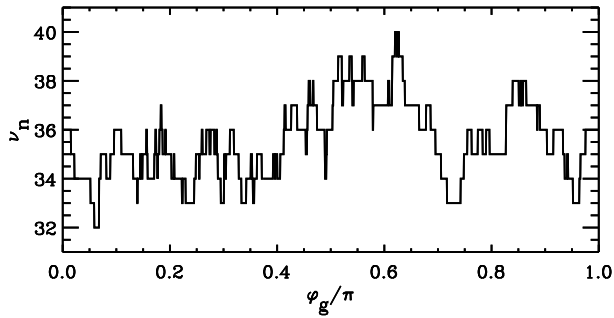}\\
  \caption{\label{fig:NodalPhi2} Number of nodal
    domains $\nu_n$ as a function of a global phase $\varphi_g$ for a wavefunction at $n_{\rm{Weyl}} \approx$ 229.}
\end{figure}

Because the phase rigidity for this wavefunction is quite large ($|\rho|^2$=0.81) the modulus of the imaginary part $|\psi_I|$ is much smaller than the modulus of the real part $|\psi_R|$. Therefore a small change in the global phase  $\varphi_g\mapsto \varphi_g +d\varphi$ for $\varphi_g$ close to zero (or close to $\pi$) has only a weak effect on the real part of the wavefunction and on its nodal pattern. Here the number of nodal domains $\nu_n$ is less likely to change as a function of phase. On the other side for $\varphi_g$ close to $\pi/2$, there is an increased probability for changes in the nodal pattern. Indeed, if the global phase approaches $\pi/2$, changes of $\nu_n$ become more frequent. In Fig.~\ref{fig:NodalPhi2} the number of nodal domains as a function of the global phase is shown for wavefunctions at Weyl numbers $n_{\rm{Weyl}}$=229 with $|\rho|^2$=0.04. In this case the phase rigidity is small, i.\,e.\ real and imaginary parts are of the same order. Therefore already for small global phases the number of nodal domains changes and a suppression of changes in $\nu_n$ for small phases is not observed -- instead the changes occur quite uniformly over the complete range of global phases. This already shows that the phase rigidity can be observed from the number of nodal domains, as a function of the global phase, at least qualitatively.

\begin{figure}
  \includegraphics[width=8cm]{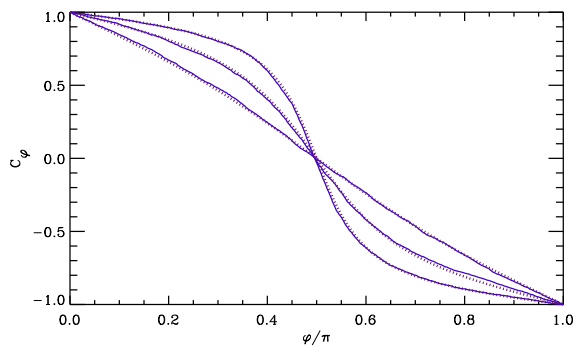}\\
  \caption{\label{fig:CorrNodal} (color online) Correlation of the
    signed areas as a function of the global phase $\varphi_g$ for three different wavefunctions at Weyl numbers $n_{\rm{Weyl}} \approx $ 229, 222, and 223 with rigidities $\rho$=0.04, 0.48, and 0.81, respectively. The curve with the smallest curvature belong to $\rho$ =0.04, and the one with the largest curvature belong to $\rho$ =0.81. The dotted lines show the results of Eq.~(\ref{eq:Corr}). The first and the last correlation corresponds to the wavefunctions used to produce Figs.~\ref{fig:NodalPhi} and \ref{fig:NodalPhi2}.}
\end{figure}

It is a non-trivial problem to obtain an expression for the number of nodal domains as a function of the global phase based on the random-wave (or the percolation) model. However there are other quantities which describe the nodal pattern as a function of the global phase and which are less resistent to a theoretical approach. Any of these is suited to exhibit the interplay between openness and nodal domains. One possibility is the global phase autocorrelation of the signed areas of the wavefunctions defined by
\begin{equation}
  \label{eq:CorrDef} C(\varphi)= \langle
  \rm{sgn}(\psi_{R,\varphi_1}) \cdot \rm{sgn}(\psi_{R,\varphi_2})\rangle, \qquad
  \varphi=\varphi_2-\varphi_1,
\end{equation}
where $\mathrm{sgn}(x)$ denotes the sign of $x$, and $\langle \cdot\rangle$ denote an average over the area. $\psi_{R,\varphi}$ denotes the real part of the wavefunction for the value $\varphi$ of the global phase. $\varphi_1$ is chosen to be the global phase, where the real and imaginary parts are uncorrelated, in our case $\varphi_1$ = 0. Equivalently one can write
\begin{equation}
  \label{eq:CorrDef2} C(\varphi)=\frac{A_{\vartheta,+}-A_{\vartheta,-}}{
    A}=
  1-2\frac{A_{\vartheta,-}}{A}
\end{equation}
where $A$ is the area of the billiard, $A_{\vartheta,+}$ is the area where $\psi_{R,\varphi_1}$ and $\psi_{R,\varphi_1}$ have the same sign, and  $A_{\vartheta,-}=A-A_{\vartheta,+}$ is the area where they have opposite signs. Thus the signed area correlator measures the fraction of the billiard area which changes sign as $\vartheta$ increases.

We can give an expression for the autocorrelation function $C(\varphi)$ based on the random wave model. At a given point we may write the wavefunction at $\varphi=0$ as
\begin{equation}
  \psi_0= \psi_R+i \psi_I= x_1 + i \xi x_2
  \label{eq:rwm_c}
\end{equation}
where $x_1$ and $x_2$ are independent and equally distributed Gaussian random variables, and $\xi$ is a real constant, that is related to the rigidity via
\begin{equation}
  \rho=\frac{\langle \psi_R^2\rangle -\langle \psi_I^2\rangle}{
    \langle \psi_R^2\rangle
    +\langle \psi_I^2\rangle}=\frac{1-\xi}{1+\xi}\ .
\end{equation}
In general, random wave models of the given type only define the rigidity on the mean over all realizations (neither the norm of the real part nor the norm of the imaginary part
are fixed). For the present the fluctuations in $\psi_R^2$ and $\psi_I^2$ just reflect the fact that we describe one point in a wavefunction while the rigidity is obtained by an integral (space average) over the billiard. Other applications of random wave models may need to take more care and define a model where the rigidity does not fluctuate from one realization to another.

Within the above variant of the random wave model the autocorrelator can easily be calculated as
\begin{eqnarray}
  \label{eq:Corr}
    C(\varphi)&=&\frac{1}{\pi}\int\ dx_1dx_2\ e^{-x_1^2-x_2^2}\times\nonumber\\
    &&\times \mathrm{sgn}\big( x_1 \big) \mathrm{sgn}\big(x_1\cos(\varphi)+\xi x_2 \sin(\varphi) \big)\nonumber\\
    &=&\int_0^{2\pi}\frac{d\alpha}{2\pi} \mathrm{sgn}\big(\cos{\alpha}\big)\mathrm{sgn}\big(\cos{\alpha}\cos{\varphi}+\xi\sin{\alpha}\sin{\varphi}\big)\nonumber\\
    &=&\frac{2}{\pi}\arctan\left(\sqrt{\frac{1+\rho}{1-\rho}}\big/ \tan{\varphi}\right)\ .
\end{eqnarray}

In Fig.~\ref{fig:CorrNodal} the global phase autocorrelation of the signed areas is shown for three different wavefunctions at approximately the same Weyl number $n_{\rm{Weyl}} \approx $ 229, 222, and 223 but for very different phase rigidities $|\rho|^2$=0.04, 0.48, and 0.810, respectively. Additionally, the results from Eq.~(\ref{eq:Corr}) are plotted as dotted lines. An excellent agreement between experiment and theory is found, especially as there is no free parameter, since the only parameter, the phase rigidity, has been determined directly from the wavefunctions.

\section{Nodal domains for the vorticity}

\begin{figure}
  \includegraphics[width=\columnwidth]{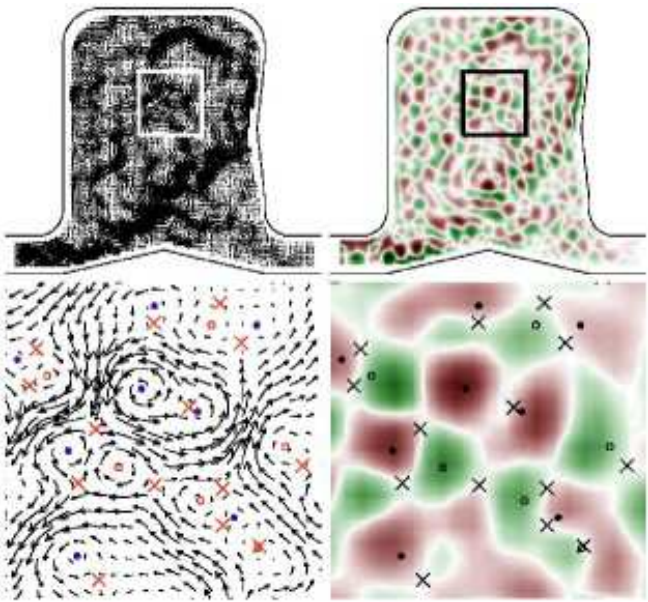}
  \caption{\label{fig:FlowVorticity} (color online) Left the
    probability current density $\vec{j}$ is shown for the whole billiard at 13.84\,GHz. On the right side the corresponding vorticity $\omega$ is plotted. The lower part of the figure shows a magnification of the region marked by squares in the upper figure. In the zoomed figures vortices and saddles are marked by circles and crosses. Open circle denote clockwise and filled circles denote counterclockwise vortices.}
\end{figure}

Since the billiard is open there is a flow through the system which, in the electromagnetic case, is described by the Poynting vector \cite{seb99}. Quantum mechanically it corresponds to the probability current density and is given by
\begin{equation}\label{eq:current}
  j \propto {\rm Im}\left[\psi^*(r)\nabla \psi(r)\right]\ .
\end{equation}

In Fig.~\ref{fig:FlowVorticity} (left) the probability current density $j$ is shown for the same wavefunction as used in Fig.~\ref{fig:NodalPhi}. A complex flow structure is observed with numerous elliptic fixed points (vortices) and hyperbolic points (saddles). The vortices and saddles are marked by circles and crosses in the lower part of the figure.

Another quantity of interest is the vorticity given by
\begin{equation}\label{eq:vorticity}
  \omega =(\nabla_x\psi_R)(\nabla_y\psi_I) - (\nabla_y\psi_R)(\nabla_x\psi_I).
\end{equation}
which, up to a constant factor, is just the curl of flow. It is shown Fig.~\ref{fig:FlowVorticity} on the right hand side. For more details on distributions and correlations of the probability current or vorticity we refer to Refs.~\onlinecite{bar02,kim03b}. One observes clearly a nodal line pattern for the vorticity. The nodal lines of the vorticity correspond to the unstable manifolds of the dynamics. Accordingly, nodal lines of the vorticity will always cross exactly at the saddle points and each vortex has a separated nodal domain with an area depending on the vortex strength and the distance to other vortices and saddles. A type of irregular checkerboard pattern of two saddles and two vortices, one clockwise, the other one counterclockwise, with a nodal line crossing between the four of them is created and can be seen in Fig.~\ref{fig:FlowVorticity}\,(upper right). Due to measurement errors but also already due to the discretisation and the bilinear interpolation these crossings will be turned into avoided crossings, thus connecting the nodal domains more or less randomly. This corresponds to the spirit of percolation theory.

\begin{figure}
  \includegraphics[width=8cm]{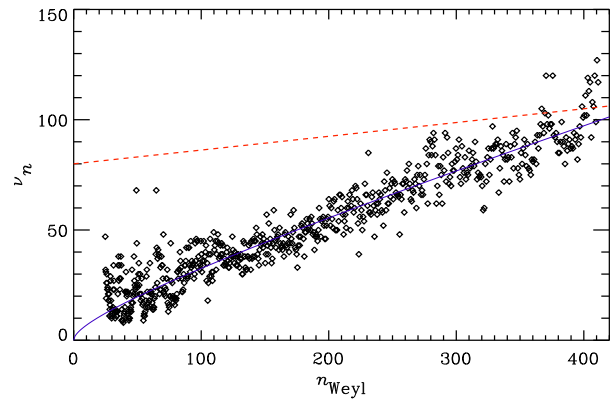}\\
  \caption{\label{fig:wnnodal} (color online) The number of nodal
    domains $\nu_n$ versus the Weyl number $n_{\rm{Weyl}}$ for the
    vorticity $\omega$ (diamonds). The red dashed line
    correspond to the theoretical prediction $\nu_n=0.0624 n$
    of the percolation model
    \cite{bog02b}. The blue solid lines are fits including boundary
    effects \cite{blu02} with Eq.~(\ref{eq:NodalDomainNumber}) with a linear slope $a$ = 0.161.}
\end{figure}

In Fig.~\ref{fig:wnnodal} we present the number of nodal domains for the vorticity. A fit with Eq.~(\ref{eq:NodalDomainNumber}) yields a linear slope of $a$=0.161 and $b$=1.65. Therefore we find again a dependency like in Eq.~(\ref{eq:NodalDomainNumber}). The slope is larger than the slope expected in the case of real wavefunctions (0.0602). Predictions for the linear increase of the vorticity in terms of a percolation model are still missing.

\section{summary}
To summarize, we have shown in this paper that the nodal domains of the real and imaginary part of the wavefunctions in open systems behave like the nodal domains for the wavefunction in closed systems. Additionally we have calculated the global phase autocorrelation of the signed areas as a function of the global phase. This quantity was shown to be an indicator of openness, i.\,e., the phase rigidity. An effect of the rigidity is also present in the fluctuations of the number of nodal domains as the global phase is varied, though an analytical description is beyond present knowledge. Nodal domains in open systems can also be defined for other quantities like the vorticity, which we have shown here. They also seem to have a linear behavior (plus square root corrections) but with a different slope in comparison to the number of nodal domains in wavefunctions in closed systems.

\section*{Acknowledgments}
The experiments were supported by the DFG.
SG thanks for the support by a research grant from the  GIF (grant I-808-228.14/2003). L.~Sirko, Warsawa, is thanked for making the geometrical data of the billiards used in Ref.~\onlinecite{Sav04b,Hul05b} available to us.

\bibliographystyle{apsrev}
\bibliography{thesis,paperdef,paper,newpaper,book}

\end{document}